\documentclass[12pt]{article}

\usepackage{sbc-template}

\usepackage{graphicx,url}
\usepackage{amsfonts}
\usepackage{amsmath}
\usepackage{amssymb}
\usepackage{dsfont}
\usepackage{color}
\usepackage[latin1]{inputenc}

\title{A Type System for Parallel Components}

\author{Francisco Heron de Carvalho-Junior\inst{1}, Rafael Dueire Lins\inst{2}}

\address{Mestrado e Doutorado em Ciência da Computa\c{c}\~ao -- Universidade Federal do Cear\'a\\
  Av. Mister Hull, s/n -- 60.455-760 -- Fortaleza/CE, Brazil
\nextinstitute
  Departamento de Eletr\^onica e Sistemas -- Universidade Federal de Pernambuco\\
  Recife/PE, Brazil
  \email{heron@lia.ufc.br, rdl@ufpe.br}
}

\begin{document}

\maketitle

\begin{abstract}

  The \# component model was proposed to improve the practice of parallel programming. This paper introduces a type system for \# programming systems, aiming to lift the abstraction and safety of programming for parallel computing architectures by introducing a notion of abstract component based on universal and existential bounded quantification. Issues about the implementation of such type system in HPE, a \# programming system, are also discussed.

\end{abstract}

%\begin{resumo}

%\end{resumo}

\section{Introduction}

Multi-core processors have already made parallel computing a mainstream technology, but high performance computing (HPC) applications that run on clusters and grids have already attracted the investments of the software industry.
The key for reaching peak performance is the knowledge of how to apply
HPC techniques for parallel programming by looking at the particular features of the parallel computing architecture.

With the raising of complexity and scale of HPC applications \cite{Post2005a}, HPC developers now demands for software engineering artifacts to develop HPC software \cite{Sarkar2004}.
Unfortunately, parallel programming is still hard to be incorporated into
usual software development platforms \cite{Bernholdt2004}.
Due to the success of component technologies in the commercial scenario, component models and frameworks for HPC applications
have been proposed \cite{Steen2006}, such as CCA and its compliant frameworks
\cite{Armstrong2006}, Fractal/ProActive \cite{Bruneton2002},
and GCM \cite{Baude2008}.
However, the HPC community still looks for a general
notion of parallel component and better connectors for efficient parallel
synchronization.

The \# component model was proposed to meet the aims of parallel
software in HPC domain.
It provides (\#-)components with the ability to be deployed in a pool
of computing nodes of a parallel execution platform and to address
non-functional concerns.
Based on a framework architecture recently proposed \cite{Carvalho2007a},
a \# programming system based on the notion of \#-components was designed and
prototyped, called HPE (\emph{The Hash Programming Environment}).
This paper presents the design of a type system for \# programming systems,
adopted in HPE, that support a suitable notion of \emph{abstract component}
for increasing the level of abstraction of parallel programming over particular
architectures with minimal performance penalties.

Section \ref{sec:hash_model} presents the \# component model and HPE. Section \ref{sec:hcl_configuration} outlines a language for describing configurations of \#-components, whose \emph{type system} is introduced in
Section \ref{sec:type_system} and implemented Section \ref{sec:implementation}.
Section \ref{sec:conclusions} concludes this paper, outlining further works.

%\begin{figure}[h]
%\centering
%\includegraphics[width=0.45\textwidth]{figures/app_example.eps}
%\caption{Example of Parallel Program} \label{fig:app_example}

\begin{figure}
\begin{center}
\begin{tabular}{lr}
\includegraphics[width=0.5\textwidth]{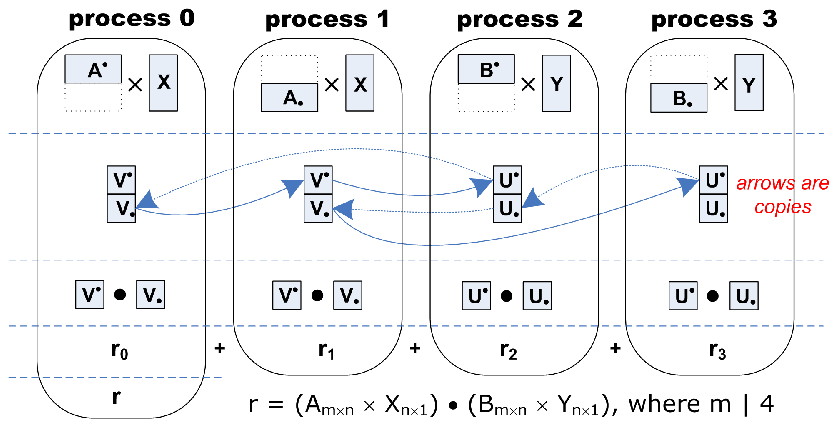} &
\includegraphics[width=0.5\textwidth]{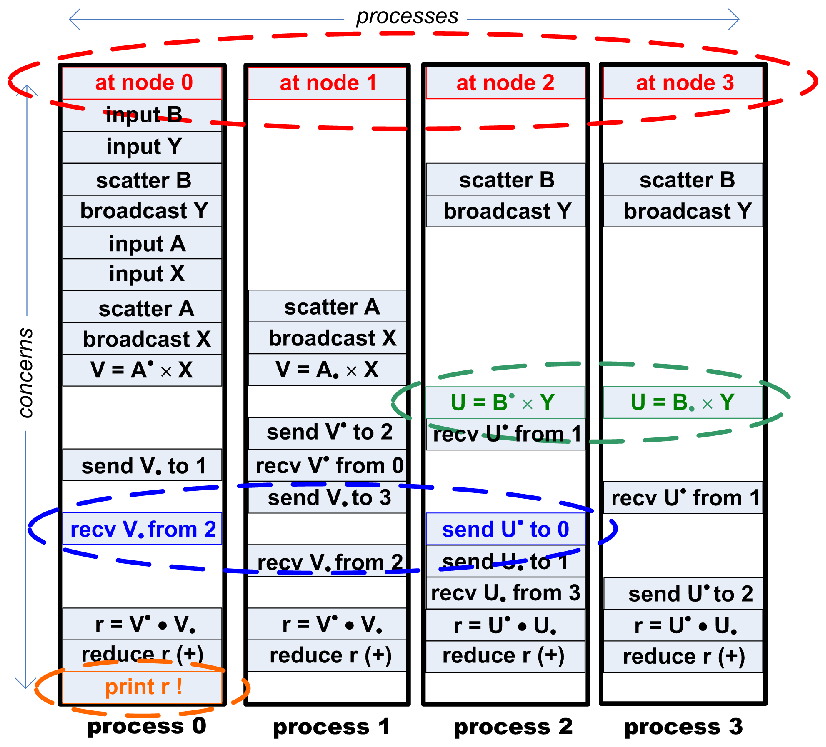}
\end{tabular}
\caption{From Processes to \#-Components, Intuitively} \label{fig:slicing_processes}
\end{center}

\end{figure}

\section{The \# Component Model}
\label{sec:hash_model}

Notions of parallel components have been proposed in many computational frameworks for
HPC applications \cite{Steen2006}.
In general, they lack the level of expressiveness and efficiency of message passing libraries such as MPI \cite{Dongarra96}. For this reason, the search for more expressive ways to express parallelism with components is at present an important research theme for people that work with CCA (Common Component Architecture), Fractal, and GCM (Grid Component Model) compliant component platforms \cite{Allan2002,Baude2007,Baduel2007}.
%In the near past, other attempts have been made with CORBA \cite{}.
The \# component model proposes a notion of components that are intrinsically parallel and shows how they can be
combined to form new components and applications.

A \emph{programming system} is defined as any artifact for development of programs for applications in some domain. Examples of programming systems are programming languages, problem solving environments, computational frameworks, visual composition languages, and so on. We say that a programming system is component-based if programs are constructed by gluing independent parts that represent some notion of \emph{component} by means of a set of supported \emph{connectors}. A component-based programming system complies to the \# component model if they support the following features:

\begin{itemize}

\item \emph{components} are built from a set of parts, called \emph{\textbf{units}}, each one supposed to be deployed in a node of a parallel computing execution platform;

\item components can be combined to form new components and applications by means of \emph{\textbf{overlapping composition}}, a kind of hierarchical composition; %whose main operations are depicted in Figure \ref{fig:overlapping_composition};

\item Each component belongs to one in a finite set of supported \emph{\textbf{component kinds}}.

\end{itemize}

Components of \# programming systems are called \#-components, which has been formally defined in previous works, using category theory and institutions \cite{Carvalho2008}.
Figure \ref{fig:slicing_processes} provides an intuitive notion of
\#-components by assuming the knowledge of the reader about the basic structure of parallel programs, as a set of processes communicating by message passing. For that, it is used a parallel program that calculate $A \times \widehat{x} \bullet B \times \widehat{y}$, where
$A_{m{\times}n}$ and $B_{m{\times}k}$ are matrices and $\widehat{x}_{n{\times}1}$ and $\widehat{y}_{k{\times}1}$ are vectors. For that, the parallel program is formed by
$N$ processes coordinated in two groups, named $p$ and $q$, with $M$ and $P$ processes, respectively. In Figure \ref{fig:slicing_processes}, $M=P=2$, $p=\{\textbf{process\ 0}, \textbf{process\ 1}\}$ and $q = \{\textbf{process\ 2}, \textbf{process\ 3}\}$.
In the first stage, the processes in $p$ calculate $\widehat{v}=A \times \widehat{x}$, while the processes in $q$ calculate $\widehat{u}=B \times \widehat{y}$, where $\widehat{v}_{m{\times}1}$ and $\widehat{u}_{m{\times}1}$ are intermediate vectors.
Figure \ref{fig:slicing_processes}(a) illustrates the partitioning of matrices and vectors and the messages exchanged (arrows).
$M^\bullet$ denotes the upper rows of the matrix $M$, where $M_\bullet$ denotes their lower rows. The definition is analogous for vectors, by taking them as matrices with a single column. Thus, the matrices $A$ and $B$ are partitioned by rows, while the vectors $\widehat{x}$ and $\widehat{y}$ are replicated across the processes in groups $p$ and $q$. After the first stage, the elements of $\widehat{v}$ and $\widehat{u}$ are distributed across the processes in groups $p$ and $q$, respectively. In the second stage, $\widehat{v}$ and $\widehat{u}$ are distributed across all the $N$ processes for improving data locality when calculating $\widehat{v} \bullet \widehat{u}$ in the third stage.

In Figure \ref{fig:slicing_processes}(b), the processes that form the parallel program described in the last paragraph are sliced according to \emph{software concern}, whose definition vary broadly in the literature \cite{Mili2004}. For the purposes of this paper, it is sufficient to take a concern as anything about the software that one wants to be able to reason about as a relatively well-defined entity. Software engineers classify concerns in functional and non-functional ones.
In the parallel program of the example, the relevant concerns include synchronization, communication and computation operations and allocation of processes onto processors. Most of them involve the participation of slices of many processes, such as the four slices that define allocation of processes to processors, the two slices of processes 2 and 3 that perform the matrix-vector product
$U = B \times Y$ in parallel,
and that ones defining communication channels (\emph{send} and \emph{recv} pairs). Such teams of cooperative slices define the \emph{\textbf{units}} of \#-components.
In Figure \ref{fig:slicing_processes}(a), candidates to be \#-components are represented by the dashed ellipses.
Thus, a \emph{unit} defines the \emph{role} of a process with respect to the \emph{concern} addressed by the \#-component. The example also shows that \#-components can deal with \emph{non-functional} concerns, such as mapping of processes onto processors. The reader may be convinced that a \# parallel programmer works at the perspective of \emph{concerns}, while a common parallel programmer works at the perspective of \emph{processes}. The resulting program may be viewed as a \#-component that encapsulates the computation of $A \times \widehat{x} \bullet B \times \widehat{y}$. In such case, the processes, numbered from 0 to 3, are their units. Notice that it is formed by combining units of the composed \#-components, taken as \emph{\textbf{slices}} of the resulting unit. This is possible due to \emph{\textbf{overlapping composition}}.
%, a kind of hierarchical composition whose formal meaning will be presented in Section \ref{sec:}.

\paragraph{Why is \# intrinsically parallel ?}

Usual component notions are sequential. In the sense of the \# component model, they are formed by only one unit. In general, parallelism is obtained by orchestration of a set of components, each one executing in different nodes. Thus, a \emph{concern} implemented in parallel must be scattered across the boundaries of a set of components, breaking encapsulation and modularization principles behind the use of components. Another common approach is to take a component as a parallel program, where parallel synchronization is introspectively implemented inside the boundaries of the parallel component using some message passing interface like MPI \cite{Dongarra96}. In such approach, the component platform is completely ``out of the way'' with communications between components and do not support hierarchical composition. Stronger parallelism approaches support parallelism by means of specific connectors for parallel synchronization, but losing flexibility and expressivity since programmers are restricted to a specific set of connectors. The scattering of implementation of components in units and the support for connectors as (\#-)components are the reasons to say that the \# programming model is intrinsically targeted at the requirements of parallel computing for high-end HPC computer architectures.

\begin{figure}
%\centering
\begin{tt}
\begin{footnotesize}

\begin{tabular}{cc}
\begin{minipage} {\textwidth}
\begin{tabbing}

$\textbf{\textit{absConfig}} \rightarrow kind\ header\ inner^*\ unit^+$\\

$header \rightarrow \textrm{configId}\ publicInner^*\ paramType^*\ cFunAppNoVar^?$ \\

$paramType \rightarrow \textrm{varId}\ cFunApp$\\

$cFunApp \rightarrow cFunAppNoVar \mid \textrm{varId}$\\

$cFunAppNoVar \rightarrow \textrm{configId}\ cFunApp^*$\\

$publicInner \rightarrow \textrm{innerId}$\\

$inner \rightarrow \textrm{innerId}\ cFunApp\ innerId^* $ \\

$unit \rightarrow \textrm{unitId}\ slice^*\ action$ \\

$slice \rightarrow \textrm{sliceId}\ \textrm{innerId}\ \textrm{unitId}$ \\

$action \rightarrow $ \\

$kind \rightarrow$ \= $\textrm{application} \mid \textrm{computation} \mid \textrm{synchronizer} \mid \textrm{data} \mid$ \\
                  \> $\textrm{environment} \mid \textrm{architecture} \mid \textrm{qualifier} $ %\\

\end{tabbing}
\end{minipage} &
\begin{minipage} {\textwidth}
\begin{tabbing}

$\textbf{\textit{concConfig}} \rightarrow$\=$header$\\
                        \>$unit^+$\\

$header \rightarrow $\=$\textrm{configId}$\\
                     \>$cFunAppNoVar$\\
                     \>$\textrm{configId}$\\
                     \>$version$ \\

$unit \rightarrow $\=$\textrm{unitId}$\\
                   \>$source$ %\\

\end{tabbing}
\end{minipage}
\end{tabular}

\end{footnotesize}
\end{tt}
\caption{HCL Abstract Syntax - Abstract ($absConfig$) and Concrete ($concConfig$)}
\label{fig:hcl_syntax}
\end{figure}

\subsection{Component Kinds}

Usual component platforms define only one general kind of component, intended to address some functional concern, with a fixed set of connectors, taken as separate entities in relation to components. The definition of component and the rules for composing them to other components define the component model of a components platform \cite{Wang2005}. It is attempted to define a notion of component that is general enough to serve for implementation of any concerns that could be encapsulated in a software module.
\# programming systems are distinct due to its support for many kinds of components, each one specialized to address specific kinds of concerns, functional or non-functional ones. We find the following main uses for components kinds:

\begin{itemize}

\item connectors are taken as specific kinds of components, making possible for a programmer to develop specific connectors for the use of their applications or libraries of connectors for reuse. This is an important feature in the context of HPC and parallel programming, where connectors must be tuned for the specific characteristics of the target parallel computer architecture.

\item component kinds can be used as an abstraction to define building blocks of applications in specific domains of computational sciences and engineering, targeting specialists from these fields. In such case, component kinds and their composition rules could be viewed as a kind of DSL (Domain Specific Language).

\item In HPC context, to ensure interoperability in the implementation of existing component-based computational frameworks is considered a hard problem. We conjecture that interoperability among many \# programming systems, specific and general purpose, may be obtained by developing of specific sets of component kinds only intended for supporting interoperability.

\end{itemize}

%\begin{figure}
%\centering
%\begin{tabular}{lr}
%\includegraphics[width=1.0\textwidth]{figures/hash_components.eps}
%\end{tabular}
%\caption{Overlapping Composition} \label{fig:overlapping_composition}
%\end{figure}

\subsection {HPE - A General Purpose \# Programming System Targeting Clusters}
\label{sec:hpe}

The \emph{Hash Programming Environment} (HPE) is a \# programming system
based on a recently proposed architecture for frameworks from which programming platforms targeting at specific application domains may be
instantiated \cite{Carvalho2007a}. It is an open-source project hosted at \texttt{\small http://code.google.com/p/hash-programmin-environment}.
The HPE framework is implemented as a plug-in to the IBM Eclipse Platform,
from which HPE is instantiated for general purpose parallel programming
of HPC applications targeting clusters of multiprocessors.
To fit this application domain, HPE supports seven kinds of components:
\emph{computations}, \emph{data structures},
\emph{synchronizers}, \emph{architectures}, \emph{environments}, \emph{applications},
and \emph{qualifiers}.
The HPE architecture has three main components:
\begin{itemize}
\item the \textsc{Front-End}, from which \emph{programmers} build configurations
of \#-components and control their life cycle;
\item the \textsc{Core}, which
manages a library of \emph{\#-components} distributed across a set of
locations and provides configuration services; and
\item the \textsc{Back-End},
which manages the \emph{components infrastructure} where \#-components are
deployed and the \emph{execution platforms} where they execute.
\end{itemize}
The interfaces between these three components were implemented as \emph{Web Services} for promoting their independence, mainly regarding localization and development platform.
For instance, from a \textsc{Front-End} a user may connect to any \textsc{Core}
and/or \textsc{Back-End} of interest that can be discovered using
UDDI services.
The \textsc{Back-End} of HPE was implemented by extending the CLI/Mono platform,
while the \textsc{Front-End} and the \textsc{Core} were implemented in Java using
the MVC (Model-View-Controller) design pattern.

\begin{figure}
%\centering
\begin{tt}
\begin{scriptsize}

\begin{tabular}{cc}

\begin{minipage}{\textwidth}
\begin{tabbing}

$\mathbf{computation}\ \textsc{M}$\=$\textsc{atVecProduct}\langle{N}\rangle(a,x,v)$\\
\> $[$\=$\textsl{T}:\textsc{Number}, \textsl{C}: \textsc{Architecture}, $\\
\> \> $\textsl{E}:\textsc{Environment}[\textsl{C}], \textsl{Da}:\textsc{MatPartition},$ \\
\> \> $\textsl{Dx}:\textsc{VecPartition},\textsl{Dv}:\textsc{VecPartition}]$ \\
$\mathbf{be}$\=$\mathbf{gin}$\\
%\> \\
\> $\mathbf{iterator}\ k\ \mathbf{from}\ 0\ \mathbf{to}\ N{-}1$\\
%\> \\
\> $\mathbf{data}\ a : \textsc{PData}\langle{N}\rangle[\textsc{Matrix}[\textsl{T}],\textsl{C},\textsl{E},\textsl{Da}]$\\
\> $\mathbf{data}\ x : \textsc{PData}\langle{N}\rangle[\textsc{Vector}[\textsl{T}],\textsl{C},\textsl{E},\textsl{Dx}]$\\
\> $\mathbf{data}\ v : \textsc{PData}\langle{N}\rangle[\textsc{Vector}[\textsl{T}],\textsl{C},\textsl{E},\textsl{Dv}]$\\
\> $\mathbf{unit}\ $\=$calculate[k]$ \\
\> $\mathbf{be}$\=$\mathbf{gin}$\\
\>            \>$\mathbf{slice}\ aslice\ \mathbf{from}\ a.matrix[k]$\\
\>            \>$\mathbf{slice}\ xslice\ \mathbf{from}\ x.vector[k]$\\
\>            \>$\mathbf{slice}\ vslice\ \mathbf{from}\ v.vector[k]$\\
\>            \>$\mathbf{action}\ \dots$\\
\> $\mathbf{end}$\\
$\mathbf{end}$

\end{tabbing}
\end{minipage}
&

\begin{minipage}{\textwidth}
\begin{tabbing}

$\mathbf{comp}$\=$\mathbf{utation}\ \textsc{MatVecProductImplForDouble}\langle{N}\rangle$\\
               \>$\mathbf{implements}\ \textsc{Mat}$\=$\textsc{VecProduct}\langle{N}\rangle$\\
               \> \> $[$\=$\textsc{Double},\textsc{GNUCluster},$\\
               \> \> \> $\textsc{MPIFull}[\textsc{GNUCluster}],\textsc{ByRows},$\\
               \> \> \> $\textsc{Replicate},\textsc{Replicate}]$ \\
               \>$\mathbf{version}\ 2.2.2.1$\\
$\mathbf{be}$\=$\mathbf{gin}$\\
\> $\mathbf{iterator}\ k\ \mathbf{from}\ 0\ \mathbf{to}\ N{-}1$\\
\> $\mathbf{unit}\ calculate[k]$ \\
\> $\mathbf{be}$\=$\mathbf{gin}$\\
\> \> // \emph{source code in the host language }\\
\> $\mathbf{end}$\\
$\mathbf{end}$

\end{tabbing}
\end{minipage}

\end{tabular}

\end{scriptsize}
\end{tt}
\caption{Examples of HCL Programs (Full Syntax)}
\label{fig:example_hcl_full_syntax}
\end{figure}

%\begin{figure}
%\begin{center}
%\includegraphics[width=1.0\textwidth]{figures/example_syntax_tree.eps}
%\caption{Abstract Syntax Tree for Configuration of Figure \ref{fig:example_hcl_full_syntax}} \label{fig:hcl_example_syntax_tree}
%\end{center}
%\end{figure}

\section{A Configuration Language for \# Programming Systems}
\label{sec:hcl_configuration}

Figure \ref{fig:hcl_syntax} presents the abstract syntax of an \emph{architecture description language} (ADL) for overlapping composition of \#-components, which could be adopted by a \# programming system. This language is called HCL (\emph{Hash Configuration Language}). HPE Front-End has implemented a visual variant of HCL.

In previous papers, overlapping composition has been formalized using a calculus of terms, called HOCC (\emph{Hash Overlapping Composition Calculus}) \cite{Carvalho2009a}, and theory of institutions \cite{Carvalho2008}. In this paper, HCL is adopted to provide a more intuitive description of overlapping composition, but keeping rigor.

A \emph{configuration} is a specification of a \#-component, which may be \emph{abstract} or \emph{concrete}. Conceptually, in a \#-programming system, a \#-component is synthesized at compile-time or startup-time using the configuration information, by combining software parts whose nature depends on the component kind.
A \# programming system defines a function $\mathcal{S}$ for synthesizing \#-components from configurations. $\mathcal{S}$ is applied recursively to the inner components of a configuration and combines the units of the inner components to build the units of the \#-component. In HPE, units of a \#-component are C\# classes.

Figure \ref{fig:example_hcl_full_syntax} present examples of configurations for abstract and concrete \#-components, written in the concrete syntax of HCL, augmented with support for iterators.
%In Figure \ref{fig:hcl_example_syntax_tree}, the abstract syntax tree of \textsc{MatVecProduct} is depicted, only for illustrating the meaning of the grammar at Figure \ref{fig:hcl_syntax}.
For simplicity, in the rest of the paper we refer to an abstract \#-components as an \emph{abstract component}, and we refer to a concrete \#-component simply as a \emph{\#-component}.

Conceptually, an abstract component fully specifies the concern addressed by all of its compliant \#-components. Their parameter types, delimited by square brackets, determine the context of use for which their \#-components must be specialized.
%Parameter types are references to abstract components with parameters supplied.
For example, the abstract component \textsc{MatVecProduct} encompasses all \#-components that implement a matrix-vector multiplication specialized for a given number type, execution platform architecture, parallelism enabling environment, and partition strategies of the matrix $a$ and vectors $x$ and $v$. Such context is determined by the parameter type variables $N$, $C$, $E$, $Da$, $Dx$, and $Dv$, respectively.
For instance, the \#-component specified by \textsc{MatVecProductImplForDouble} is specialized for calculations with matrices and vectors of \underline{double precision float point numbers}, using MPI for enabling parallelism, targeting a GNU Linux cluster, and supposing that matrix $a$ is partitioned by rows, and that elements of vectors $x$ and $v$ are replicated across processors. This is configured by supplying parameter type variables of \textsc{MatVecProduct} with appropriate abstract components that are subtypes of the bound associated to the supplied variable (e.g. $\textsc{Replicate} <: \textsc{VecPartition}$).

In the body of a configuration, a set of inner components are declared, whose overlapping composition form the component being configured. In \textsc{MatVecProduct}, they are identified by $a$, $x$, and $v$ and typed by a reference to a configuration of abstract component with its context parameters supplied. Indeed, the inner component $a$ is of kind \emph{data} and it is obtained from the configuration $\textsc{PData}$ when applied to the context parameters $\textsc{Matrix}[N]$, $C$, $E$, and $Da$, which means that it is a parallel matrix of numbers of some configuration abstracted in the variable $N$, partitioned using the partitioning strategy defined by the variable $Da$, specialized for the execution platform $C$, and for the parallelism enabling environment $E$. These variables come from the enclosing configuration.

The header of a configuration written in HCL also informs its \emph{kind} and a set of \emph{component parameters}, which are references to inner components defined as \emph{public} ones.
In fact, component parameters provide \emph{high-order} features for \#-components \cite{Alt2004}. In the example, all the inner components - $a$, $x$, and $v$ - must be received as parameters by \textsc{MatVecProduct} compliant \#-components in execution time.

Finally, a configuration declares a non-empty set of units, formed by \emph{folding} units of inner components, called \emph{slices} of the unit being declared. \textsc{MatVecProduct} has $N$ units named $calculate$. Their \emph{slices} define the local partitions of $a$, $x$, and $v$. In a well formed configuration, all units of any inner component are slices of some unit of the abstract component being configured. A computation unit must also declare an action that specifies the operation to be performed. Recently, we have proposed the use of \textsf{Circus} for formal specification of these actions \cite{Carvalho2009a}.

In \textsc{MatVecProductImplForDouble}, it is provided an implementation for the units of \textsc{MatVecProduct}, using the host language for programming units of \#-components of kind computation. In HPE, computations, as well the other kinds of components, are programmed in any language that has support in the CLI/Mono platform. The HPE system partially generate the code of units of abstract components and \#-components, using the translation schema that will be presented in Section \ref{sec:implementation}.

%\subsection{Overlapping Composition Semantics in HCL}

\begin{figure}
\centering
\begin{tt}
\begin{minipage} {\textwidth}
\begin{tabular}{lllr}

$\texttt{T}$ & $\rightarrow$ & $\texttt{A}^n$\ $^{n{\geq}0}$ | $\texttt{H}$ | $\mathtt{Top}_\kappa$ & \textrm{\textbf{(4.1)}} \\ \\
$\texttt{A}^n$ & $\rightarrow$ & $\left[X_i <: \texttt{H}_i\ ^{i=1{\dots}n}\right] \rhd \texttt{A}^0$\ \ \, $^{n{\geq}1}$  & \textbf{\textrm{(4.2)}} \\ \\
$\texttt{A}^0$ & $\rightarrow$ & $\kappa \bullet \left\langle{a_i:\texttt{H}_i\ ^{i=1{\dots}k}}\right\rangle \rightarrow \left\langle{a_{i}:\texttt{H}_{i}\ ^{i=k{+}1{\dots}l}}, u_i{:}\left\langle{\sigma_i, L_i}\right\rangle\ ^{i=1{\dots}q}\right\rangle$ & \textbf{\textrm{(4.3)}} \\ \\
$\texttt{H}$ & $\rightarrow$ & $X$ | $\texttt{A}^n \lhd \left[\texttt{H}_i\ ^{i=1{\dots}n}\right]$\ \ \, $^{n{\geq}1}$ & \textbf{\textrm{(4.4)}}

\end{tabular}
\end{minipage} %&
\end{tt}
\caption{Configuration Types}
\label{fig:lambda_hash_calculus_syntax}
\end{figure}

\begin{figure}[b]

\begin{tt}
\begin{scriptsize}

\[
\begin{array}{ccc}
\begin{array}{c}
\Gamma \vdash \texttt{S} <: \texttt{S}
\end{array}
&
\begin{array}{c}
\Gamma \vdash \texttt{S}{<:}\texttt{U} \hspace{0.5cm}  \Gamma \vdash \texttt{U}{<:}\texttt{T} \\
\hline
\Gamma \vdash \texttt{S}{<:}\texttt{T}
\end{array}
&
\begin{array}{c}
\mbox{\rm $\kappa$ is the kind in the shape of $\texttt{S}$}\\
\hline
\Gamma \vdash \texttt{S} <: \mathtt{Top}_\kappa
\end{array}
\\
\textrm{\it (Reflexive)} & \textrm{\it (Transitive)} & \textrm{\it (Top)}
\end{array}
\]

\[
\begin{array}{cc}
\begin{array}{c}
%\Gamma \vdash {T}_1{<:}{T}_2 \hspace{0.5cm}
\Gamma, X{<:}{T}     \vdash {S}_1{<:}{S}_2\\
\hline
\Gamma \vdash [X{<:}{T}] \rhd {S}_1 <: [X{<:}{T}] \rhd {S}_2
\end{array}
&
\begin{array}{c}
\Gamma \vdash {S}_1{<:}{S}_2,\ \ \ \Gamma \vdash {T}_2{<:}{T}_1\\
\hline
\Gamma \vdash {S}_1 \lhd [{T}_1] <: {S}_2 \lhd [{T}_2]
\end{array}
\\
\textrm{\it (Abstract Component)} & \textrm{\it (\#-Component)}
\end{array}
\]

\[
\begin{array}{c}
\{a_i^1\ ^{i=1{\dots}k} \} = \{a_i^2\ ^{i=1{\dots}l} \},\ \{a_i^1\ ^{i=1{\dots}l} \} \supseteq \{a_i^2\ ^{i=1{\dots}k} \},\ \sigma_i^1 \supseteq \sigma_i^2\ ^{i=1{\dots}q},\ L_i^1/\sigma_i^2 \subseteq L_i^2 \ ^{i=1{\dots}q} \\
\exists i,j \in \{1{\dots}k\} \mid a_i^1 = a_j^2 \Rightarrow \Gamma \vdash {T}_j^2 <: {T}_i^1,\ \exists i \in \{k{+}1{\dots}l^1\}, j \in \{k{+}1{\dots}l^2\} \mid a_i^1 = a_j^2 \Rightarrow \Gamma \vdash {T}_i^1 <: {T}_j^2\\
\hline
\Gamma \vdash \kappa \bullet \left\langle{a_i^1{:}{T}_i^1\ ^{i=1{\dots}k}}\right\rangle \rightarrow \left\langle{a_{i}^1{:}{T}_{i}^1\ ^{i=k{+}1{\dots}l^1}}, u_i{:}\left\langle{\sigma_i^1, L_i^1}\right\rangle\ ^{i=1{\dots}q}\right\rangle \\
\hspace*{5cm}<: \kappa \bullet \left\langle{a_i^2{:}{T}_i^2\ ^{i=1{\dots}k}}\right\rangle \rightarrow \left\langle{a_{i}^2{:}{T}_{i}^2\ ^{i=k{+}1{\dots}l^2}}, u_i{:}\left\langle{\sigma_i^2, L_i^2}\right\rangle\ ^{i=1{\dots}q}\right\rangle\\
\textrm{\it (Shape)}
\end{array}
\]

\end{scriptsize}
\end{tt}

\caption{Subtyping Rules}
\label{fig:subtyping_rules}

\end{figure}

%\newpage
\section{A Type System for \# Programming Systems}
\label{sec:type_system}

Figure \ref{fig:lambda_hash_calculus_syntax} presents a syntax for types of configurations of \#-components, whose associated subtyping relation is presented in Figure \ref{fig:subtyping_rules}. The production \textbf{4.1} states that a configuration may be typed as an abstract component type or a \#-component type. Also, it defines that there is a \emph{top} abstract component associated to each kind. Abstract component types are defined in \textbf{4.2}. The set of bound variables $X_1,\dots,X_n$ denote their \emph{context}. An abstract component type also specifies a shape, describing how it forms an abstract component from overlapping composition of other \#-components. The shape of an abstract component type is defined in \textbf{4.3}. The general form of \#-component types is defined in \textbf{4.4}, from an abstract component type by supplying their bound context variables.

%like in $\mathcal{C} \lhd [H_1,\dots,H_n]$, specifying the context for which a \#-component of that type is best tuned for \textbf{(4.4)}. Indeed, the \#-component is better tuned for a context where $X_1=H_1,\dots,X_n=H_n$. For that, $H_1<:S_1,\dots,X_n<:S_n$.

In the shape of a \#-component (Figure \ref{fig:lambda_hash_calculus_syntax}), $\kappa$ specifies its kind, among the kinds supported by the \# programming system. The labels $a_1,\dots,a_l$ identify inner components, with their associated \#-component types. The inner components labeled from $a_1$ to $a_k$ are the \emph{public} ones (component parameters of a configuration). The assertions $u_i{:}\left\langle{\sigma_i,L_i}\right\rangle\ ^{i=1{\dots}q}$ type the units of the \#-component.
For any unit, the function $\sigma$ maps a set of symbols that denote labels of \emph{slices} to units of inner components, denoted by $a.u$, where $a \in \{a_1,\dots,a_l\}$ and $u$ is a label of a unit of the inner component labeled by $a$. The typing rules for configurations impose that each unit of an inner component must be a slice of one, and only one, unit of the \#-component. $L$ is a formal language on the alphabet $\mathbf{Dom}(\sigma)$, denoting the tracing semantics that defines the action of the unit.
% A \# programming system may define how the trace semantics of units is expressed and translated to code.

\begin{figure}
\begin{tiny}
\begin{tabular}{c}

\makebox{
\begin{tabular}{c}
\makebox{
\begin{math}
\begin{array}{c}
v_{ij} \in \mathbf{units\_of}(U_{z_{ij}})\ ^{j=1{\dots}m_i}\ ^{i=1{\dots}l}, \\
\{r_{ij}\ ^{j=1{\dots}p_i\ {i=1{\dots}l}}\} \subseteq \{1,{\dots},l\},\ \ \ \ \ \ \{z_{ij}\ ^{{j=1{\dots}m_i}\ {i=1{\dots}l}}\} = \{1,{\dots},l\}
\end{array}
\end{math}
}
\\
\hline
\\
\makebox{
\begin{math}
\begin{array}{c}
\mathcal{T}\left(\makebox{\mbox{
\fbox{\begin{minipage} {\textwidth}
\begin{tabbing}
$\mathit{kind}\ C\ [X_i:T_i\ ^{i=1{\dots}n}](a_i\ \ ^{i=1{\dots}k})$ \\
$\mathbf{beg}$\=$\mathbf{in}$ \\
     \> $kind_i\ a_i : U_i\ (a_{r_{ij}}\ ^{j=1{\dots}p_i})\ \ ^{i=1{\dots}l}$\\
     \> $\mathbf{unit}\ u_i\ \ \ ^{i=1{\dots}q}$ \\
     \> $\mathbf{beg}$\=$\mathbf{in}$ \\
     \>      \>$\mathbf{slice}\ s_{ij}\ \mathbf{from}\ a_{z_{ij}}.v_{ij}\ \ ^{j=1{\dots}m_i}$ \\
     \>      \>$\mathbf{action}\ A_i$\\
     \> $\mathbf{end}$\\
$\mathbf{end}$
\end{tabbing}
\end{minipage}}}
}
, \Gamma
\right)
\end{array}
\end{math}
=
\makebox{
\begin{minipage}{\textwidth}
\begin{tabbing}
$[X_i{<:}$\=$\mathcal{T}\left(T_i\right)\ ^{i=1{\dots}n}]\ \rhd\ kind \bullet$ \\
          \> $\langle a_i{:}\mathcal{T}$\=$\left(U_i\left(U_{r_{ij}}\ ^{j=1{\dots}p_i}\right),\Gamma \cup \{X_i\ ^{i=1{\dots}n}\}\right)\ ^{i=1{\dots}k} \rangle$\\
          \>      \> $\rightarrow$ $\langle$\=$ a_i{:}\mathcal{T}\left(U_{i}\left(U_{r_{ij}}\ ^{j=1{\dots}p_{i}}\right),\Gamma \cup \{X_i\ ^{i=1{\dots}n}\}\right)\ ^{i=k+1{\dots}{l}},$\\
          \>            \>        \>                     $u_i{:}\left\langle{\{s_{ij} \mapsto a_{z_{ij}}.v_{ij}\ ^{j=1{\dots}m_i}\}, \mathcal{L}(A_i)}\right\rangle\ ^{i=1{\dots}q}\rangle$%\\

\end{tabbing}
\end{minipage}
}
}
\end{tabular}
}
\\
\textbf{\footnotesize (5.1)}
\\ \\ \\
\makebox{
\begin{tabular}{c}
$\mathcal{T}(V_i,\Gamma) <: \mathcal{T}(T_{i},\Gamma)\ ^{i=1{\dots}n}$\\
\hline\\
\makebox{
\begin{math}
\begin{array}{c}
\mathcal{T}\left(\makebox{\mbox{
\fbox{\begin{minipage} {\textwidth}
\begin{tabbing}
$\mathit{kind}\ C\ [X_i:T_i\ ^{i=1{\dots}n}](\mathrm{A})$ \\
$\mathbf{begin}\ Body\ \mathbf{end}$
\end{tabbing}
\end{minipage}}}
} \left[V_i\ ^{i={1{\dots}n}}\right],\Gamma \right)
\end{array}
\end{math}
=
\begin{math}
\begin{array}{c}
\mathcal{T}\left(\makebox{\mbox{
\fbox{\begin{minipage} {\textwidth}
\begin{tabbing}
$\mathit{kind}\ C\ [X_i:T_i\ ^{i=1{\dots}n}](\mathrm{A})$ \\
$\mathbf{begin}\ Body\ \mathbf{end}$
\end{tabbing}
\end{minipage}}}
}, \emptyset
\right) \lhd \left[\mathcal{T}\left(V_i,\Gamma\right)\ ^{i=1{\dots}n}\right]
\end{array}
\end{math}
}
\end{tabular}
}\\
%\\ \\
\\
\textbf{\footnotesize (5.2)}
\\ \\
\\
\makebox{
\begin{tabular}{c}
$\mathcal{T}(B_i,\Gamma) <: \mathcal{T}(U_{i},\Gamma)\ ^{i=1{\dots}k}$\\
\hline\\
\makebox{
\begin{math}
\begin{array}{c}
\mathcal{T}\left(\makebox{\mbox{
\fbox{\begin{minipage} {\textwidth}
\begin{tabbing}
$\mathit{kind}\ C\ [X_i:T_i\ ^{i=1{\dots}n}](a_i\ \ ^{i=1{\dots}k})$ \\
$\mathbf{beg}$\=$\mathbf{in}$ \\
     \> $kind_i\ a_i : U_i\ (a_{r_{ij}}\ ^{j=1{\dots}p_i})\ \ ^{i=1{\dots}l}$\\
     \> $\mathbf{unit}\ u_i\ \ \ ^{i=1{\dots}q}$ \\
     \> $\mathbf{beg}$\=$\mathbf{in}$ \\
     \>      \>$\mathbf{slice}\ s_{ij}\ \mathbf{from}\ a_{z_{ij}}.v_{ij}\ \ ^{j=1{\dots}m_i}$ \\
     \>      \>$\mathbf{action}\ A_i$\\
     \> $\mathbf{end}$\\
$\mathbf{end}$
\end{tabbing}
\end{minipage}}}
} \left[\mathrm{\normalsize V}\right]\ \left(B_i\ ^{i={1{\dots}k}}\right)
,\Gamma \right)
\end{array}
\end{math}
=
\begin{math}
\begin{array}{c}
\mathcal{T}\left(\makebox{\mbox{
\fbox{\begin{minipage} {\textwidth}
\begin{tabbing}
$\mathit{kind}\ C\ [X_i:T_i\ ^{i=1{\dots}n}]$ \\
$\mathbf{beg}$\=$\mathbf{in}$ \\
     \> $kind_i\ a_i : B_i\ (a_{r_{ij}}\ ^{j=1{\dots}p_i})\ \ ^{i=1{\dots}k}$\\
     \> $kind_i\ a_i : U_i\ (a_{r_{ij}}\ ^{j=1{\dots}p_i})\ \ ^{i=k+1{\dots}l}$\\
     \> $\mathbf{unit}\ u_i\ \ \ ^{i=1{\dots}q}$ \\
     \> $\mathbf{beg}$\=$\mathbf{in}$ \\
     \>      \>$\mathbf{slice}\ s_{ij}\ \mathbf{from}\ a_{z_{ij}}.v_{ij}\ \ ^{j=1{\dots}m_i}$ \\
     \>      \>$\mathbf{action}\ A_i$\\
     \> $\mathbf{end}$\\
$\mathbf{end}$
\end{tabbing}
\end{minipage}}}
} \left[\mathrm{\normalsize V}\right]
,\Gamma \right)
%\lhd \left(\mathcal{T}\left(B_i\right)\ ^{i=1{\dots}k}\right)
\end{array}
\end{math}
}
\end{tabular}
}
\\
\textbf{\footnotesize (5.3)}
\\ \\
\makebox{
\begin{tabular}{cc}
\begin{tabular}{c}
$X \in \Gamma$\\
\hline
$\mathcal{T}(X,\Gamma)=X$ \\
\\
\textbf{\footnotesize (5.4)}
\end{tabular}
&
\begin{tabular}{c}
\makebox{
\begin{math}
\mathcal{T}\left(\makebox{\mbox{
\fbox{\begin{minipage} {\textwidth}
\begin{tabbing}
$\mathit{kind}\ $\=$C\ [X_i:T_i\ \ ^{i=1{\dots}n}]$\\
                 \>$\mathbf{implements}\ C'\ [X_i\ ^{i=1{\dots}n}]$ \\
$\mathbf{beg}$\=$\mathbf{in}$ \\
     \> $\mathbf{unit}\ u_i\ \ \ ^{i=1{\dots}q}$ \\
     \> $\mathbf{beg}$\=$\mathbf{in}$ \\
     \>      \>$\emph{\textbf{S}}_i$\\
     \> $\mathbf{end}$\\
$\mathbf{end}$
\end{tabbing}
\end{minipage}}}
}, \Gamma
\right)
\end{math}
}
=
\makebox{
\begin{minipage}{\textwidth}
\begin{tabbing}

$\mathcal{T}\left({C'}\ \left[T_i\ ^{i=1{\dots}n}\right], \Gamma \right)$
%$\mathcal{T}\left({C'}\right) \lhd \left[\mathcal{T}\left(T_i\right)\ ^{i=1{\dots}n}\right]$

\end{tabbing}
\end{minipage}
}\\
\\
\textbf{\footnotesize (5.5)}
\end{tabular}
\end{tabular}
}

\end{tabular}
\end{tiny}
\caption{Schema $\mathcal{T}$ for Computing the Type of a Configuration (Outline)}
\label{fig:configuration_typing}
\end{figure}

In Figure \ref{fig:configuration_typing}, it is outlined $\mathcal{T}$, a function for calculating the type of a configuration. The auxiliary parameter $\Gamma$ is the set of bound variables, often known as context. It ensures that any variable referred in a configuration is declared in the header. No free variables exist in a well formed configuration.
$T$, $U$, and $V$ denote logical variables in the definition for references to configurations of \#-components. The definition \textbf{5.1} types the configuration of an abstract component. The definition \textbf{5.2} types a configuration of abstract component applied to an actual context, where the resulting type is the type of the \#-component that may be applied in the context. The definition \textbf{5.3} types an abstract component with public inner components supplied, which is necessary to define the type of inner components in the definition \textbf{5.1}. The definition \textbf{5.4} only maps configuration variables to type variables, provided that they exist in the context. The definition \textbf{5.5} types a configuration of a \#-component. For simplicity, the definition ignores \textbf{extends} clauses of configurations (definition by inheritance).

%Indeed, it specifies that a configuration defines a \#-component of an \emph{abstract component type} or \emph{concrete component type}.

\subsection{Interpretation}

Abstract and concrete components may be interpreted in terms of the combinators of an usual type system with universal and existential bounded quantification and type operators. Let $\mathbf{C}$ be an abstract component type $\mathbf{C} \equiv [Z <: \texttt{T}_1] \rhd \texttt{T}_2$. Its interpretation, $\mathbf{C}^{\mathcal{I}}$ may be defined like below:

\begin{minipage}{\textwidth}
\centering

$\mathbf{C}^{\mathcal{I}} \equiv \lambda{X}{<:}{T}_1.\ \forall{Y}{<:}X.\ \{\exists{Z}{<:}{T}_1; {T}_2\}$,

\end{minipage}

where variables $X$ and $Y$ are not referenced in ${T}_2$ $\left(\{X,Y\} \cap \mathit{Vars}({T}_2) = \emptyset\right)$.
Moreover, a \#-component
\[
\emph{kind}\ \mathbf{c}\ [Y{:}{T}_3]\ \mathbf{implements}\ \mathbf{C}\ [Y]\ \mathbf{begin}\ \dots\ \mathbf{end}\\
\]
has the following interpretation:

\begin{minipage}{\textwidth}
\centering

$\mathbf{c}^{\mathcal{I}} \equiv \lambda{Y}{<:}{T}_3.\ \left(\{*Y;{t}\}\ \mathbf{as}\ \{\exists{Z}{<:}{T}_1; \texttt{T}_2\}\right)$

\end{minipage}

\begin{figure}
\centering
\begin{tabular}{c|c}

\begin{minipage}{\textwidth}
\begin{scriptsize}
\begin{tabbing}

\textbf{proc}\=\textbf{edure} \emph{\underline{findHashComponent}}\\
    \> \emph{\underline{sort}}(\textsc{CTop});\\
    \> \emph{\underline{tryGeneralize}}(next of \textsc{CTop});\\
    \> \textbf{if} (\= \textsc{cTop} has implementation) \\
    \>     \> \textbf{then} \textbf{return} implementation of \textsc{CTop}\\
    \>     \> \textbf{else} fail !\\
    \> \textbf{end-if}\\
\textbf{end-procedure}\\
\\
\textsf{lastMarked} = null;\\
\textbf{proc}\=\textbf{edure} \emph{sort}($C$): \\
\> \textbf{if} \= $C$ has unmarked parameters \\
\>	  \> \textbf{for} \= \textbf{each} unmarked parameter $C_i$ \textbf{of} $C$  \\
\>	  \>     \> \emph{\underline{sort}}($C_i$)\\
\> \textbf{else} \= \\
\>      \>   mark($C$);\\
\>      \>   next of $C$ = \textsf{lastMarked};\\
\>      \>   \textsf{lastMarked} = $C$;\\
\> \textbf{end-if}  \\
\textbf{end-procedure}

\end{tabbing}
\end{scriptsize}
\end{minipage}

&

\begin{minipage}{\textwidth}
\begin{scriptsize}

\begin{tabbing}
\emph{assumption}: let $\kappa$ be the kind of \textsc{CTop}.\\
\\
\textbf{proc}\=\textbf{edure} \emph{\underline{tryGeneralize}}($C$) \\
    \> \textbf{if} \=$C$=null or \textsc{CTop} has not an implementation in $\epsilon$ \\
    \>    \> reset $C$;\\
	\>    \> $C'$ = $C$;\\
	\>    \> \textbf{rep}\=\textbf{eat} \\
	\>    \>    \> replace $C$ by $C'$\\
    \>    \>    \> \emph{\underline{tryGeneralize}}(next of $C$)\\
	\>    \>    \> $C'$ $\leftarrow$ least proper supertype of $C$;\\
	\>    \> \textbf{until} $C'$ == $\mathbf{Top}_\kappa$ \textbf{or} \textsc{CTop} has an implementation in $\epsilon$;\\
    \> \textbf{end-if}\\
\textbf{end-procedure}

\end{tabbing}
\end{scriptsize}
\end{minipage}

%\\
%(a) & (b)

\end{tabular}

\vspace{0.5cm}

\begin{scriptsize}
\begin{minipage}{\textwidth}
\textsc{CTop} is a \emph{\#-component type}. Thus, it has form $\mathbf{H}$, such that $\mathbf{H} \equiv \textit{cid}\ [C_1,C_2,\dots,C_n]$, where $\textit{cid}$ is a reference to a configuration of abstract component and each $C_i$ is a context parameter of the form $\mathbf{H}$, recursively. The \emph{resolution algorithm} tries to find a \#-component that types to \textsc{CTop} in an environment $\epsilon$ of deployed \#-components maintained by the \# programming system.
The algorithm has two phases, defined by the procedures \emph{sort} and \emph{tryGeneralize}.
The first one calculates a total order for traversing
the recursive context parameters of \textsc{CTop}, by calculating the relation ``next of''.
Procedure \emph{tryGeneralize} recursively traverse this list, calculating the least proper supertype of each parameter in $\epsilon$ and testing if
the current generalized type has some implementation in the environment $\epsilon$.
If anyone is found, the procedure returns it.
The operation ``replace $C$ by $C'$'' replaces, in \textsc{CTop}, the parameter $C$ by its least supertype $C'$ in $\epsilon$, while ``reset $C$''
sets $C$ back to the initial parameter, after successive generalizations.
\textbf{The algorithm always stop}, since there is a finite number of parameters in an abstract component and each kind $\kappa$ of abstract component
has a maximum supertype ($\mathtt{Top}_\kappa$). Also, \textbf{the algorithm is deterministic}, because each abstract component has
only one supertype (by single inheritance) and each abstract
component has only one \#-component that conforms to it in the context (by singleton design pattern).
\end{minipage}
\end{scriptsize}
\caption{Deterministic Traversing of Subtypes of Abstract Components}
\label{fig:generalization_algorithm}
\end{figure}

Notice that
$\mathbf{c}^{\mathcal{I}}$ has type $\mathbf{C}^{\mathcal{I}}\ [{T}_3]$\footnote{Using the notation of \cite{Pierce2002}. $\mathbf{C}^{\mathcal{I}}$ is a \emph{type operator}, with \emph{parameter type} $X$ bounded by $T_1$. Thus, the type $\mathbf{C}^{\mathcal{I}}[\mathrm{U}]$, for a given $\mathrm{U}{<:}T_1$, is universally quantified (\emph{polymorphism}) in the variable $Y$. A \#-component $\mathbf{c}^{\mathcal{I}}$ applied to $\mathrm{V}$ ($\mathbf{c}^{\mathcal{I}}\ \mathrm{V}$), typed as $\mathbf{C}^{\mathcal{I}}[\mathrm{U}]$, where $\mathrm{V}{<:}\mathrm{U}$, returns a \emph{package} of \emph{existential type} $\{\exists{Z}{<:}{T}_1; {T}_2\}$ with an abstract \emph{representation type} bounded by $V$, which is safe since $\mathrm{V}{<:}\mathrm{U}{<:}T_1$.}.

As discussed before, the declaration of an inner component of abstract component type $\mathbf{C}$ makes explicit the definition of the intended context in the supplied parameters of $\mathbf{C}$.
For instance, suppose that $\mathbf{c}$ is dynamically linked by the execution environment for supplying the inner component labeled $a$, defined as

\begin{minipage}{\textwidth}
\centering

$\mathbf{\mathit{kind}}\ a : \mathbf{C}\ [{T}_3']$.

\end{minipage}

Of course, $T_3' <: T_3$. Thus, ${T}_3'$ is now the representation type in $\mathbf{c}$, which has been generically defined as $X$, such that $X{<:}{T}_3$. In terms of the interpretation, it is applied the package $\mathbf{c}^{\mathcal{I}}\ T_3'$ in the context.
All operations inside $\mathbf{c}$ will be defined in relation to ${T}_3'$ and not in relation to ${T}_3$, the upper bound of the abstract representation type $X$.

More intuitively, $\mathbf{C}$ includes \#-components that are best tuned to be applied in a context where a subtype of $T_1$ is used, abstracted in type variable operator $X$. In particular, the previous context, for inner component $a$, requires $T_3'$. Thus, any \#-component belonging to $\mathbf{C}$ that is best tuned for some supertype of $T_3'$ and subtype of $T_1$, may be dynamically bound to $a$, such as $\mathbf{c}$, which best tuned for $T_3$, since $T_3'{<:}T_3{<:}T_1$.

The previous discussion may be trivially generalized for many parameters.

For improving understanding, let

\begin{center}
\begin{footnotesize}
\begin{minipage}{\textwidth}
\centering
\begin{tabbing}
   $\mathbf{synchronizer}\ \textsc{Channel}\ [E{:}\textsc{Environment}, D{:}\textsc{Data}]$\\
%   $\vdots$\\
   $\mathbf{beg}$\=$\mathbf{in}$ \\
                 \>$\mathbf{unit}\ send\ $\\%\mathbf{begin} \dots \mathbf{end}$\\
                 \>$\mathbf{unit}\ recv\ $\\%\mathbf{begin} \dots \mathbf{end}$\\
   $\mathbf{end}$
\end{tabbing}
\end{minipage}
\end{footnotesize}
\end{center}

be a configuration of an abstract component whose \#-components represent communication channels that may be tuned for a specific parallelism enabling \emph{environment} (middleware or library) and \emph{data type} to be transmitted. A configuration may demand for
\begin{center}
\begin{minipage}{\textwidth}
\centering
$\mathbf{synchronizer}\ {ch} : \textsc{Channel}\ [\textsc{MPIFull}, \textsc{Vector}]$
\end{minipage}
\end{center}, where $ch$ must be dynamically bound to the best communication channel available in the environment that can transmit an array in an execution platform where full MPI is available (any \#-component package whose type is a subtype of $\mathcal{T}(\textsc{Channel}) \lhd [\textsc{MPIFull}, \textsc{Vector}]$). By the subtying rules, \#-components with the following configuration headers may be bound to $ch$:

\begin{footnotesize}
\begin{enumerate}

\item $\mathbf{synchronizer}\ \textsc{ChannelImpl1}\ [E{:}\textsc{MPIFull}, D{:}\textsc{Vector}]\ \mathbf{implements}\ \textsc{Channel}\ [E, D]$

\item $\mathbf{synchronizer}\ \textsc{ChannelImpl2}\ [E{:}\textsc{MPIBasic}, D{:}\textsc{Vector}]\ \mathbf{implements}\ \textsc{Channel}\ [E, D]$

\item $\mathbf{synchronizer}\ \textsc{ChannelImpl3}\ [E{:}\textsc{MPIFull}, D{:}\textsc{Data}]\ \mathbf{implements}\ \textsc{Channel}\ [E, D]$

\item $\mathbf{synchronizer}\ \textsc{ChannelImpl4}\ [E{:}\textsc{MPIBasic}, D{:}\textsc{Data}]\ \mathbf{implements}\ \textsc{Channel}\ [E, D]$

\end{enumerate}
\end{footnotesize}

The first one is the better tuned one for the context where $\mathcal{T}(\textsc{Channel}) \lhd [\textsc{MPIFull}, \textsc{Vector}]$ is demanded. The other ones are approximations. By looking at the fourth case, notice that a channel that use the basic subset of MPI primitives and that can transmit any data structure, including arrays, can be applied in the context. In fact, if the system does not find a better tuned \#-component, it will traverse subtypes, deterministically, using the algorithm described in Figure \ref{fig:generalization_algorithm}, to find a the best approximation available in the environment of deployed \#-components. In the example, by supposing that $\textsc{MPIFull}$ directly extends $\textsc{MPIBasic}$ and that $\textsc{Vector}$ directly extends $\textsc{Data}$, the types will be traversed in the presented order. To be deterministic, the so called \emph{resolution algorithm} supposes that the \# programming system supports a \emph{nominal} and \emph{single inheritance} subtyping system. In fact, both restrictions are supported by HCL.

%\subsection{Resolution Algorithm}

%From the usual typing and sub-typing rules for fully bound quantification (universal and existential), it is possible to infer typing rules $\textrm{\it T-Pack}$ and $\textrm{\it T-Uses}$ and sub-typing rules $\textrm{\it S-AbstractLeft}$ and $\textrm{\it S-AbstractRight}$.

\begin{figure}
\centering
\includegraphics[width=140.0mm]{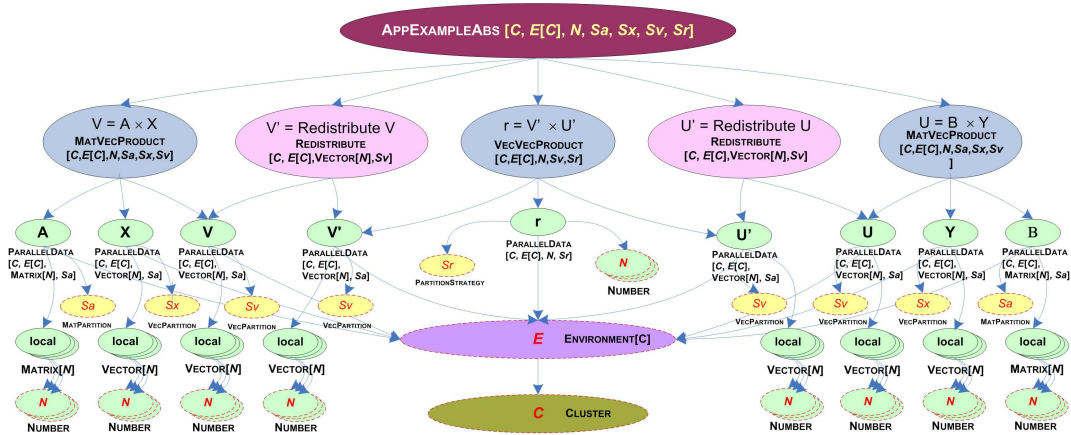}
\caption{Hierarchy of Components of the Application Example} \label{fig:hierarchy_application_example}
\end{figure}

\section{Implementation Issues}
\label{sec:implementation}

This section shows how the proposed type system has been implemented in HPE, the \# programming system introduced in Section \ref{sec:hpe}. The Back-End of HPE treats a \#-component as a set of CLI/Mono object, each one associated to a unit, instantiated from a C\# class. Therefore, the function $\mathcal{S}$ map configurations of abstract components to C\# \emph{interfaces} and configurations of \#-components to C\# \emph{classes} that implement the \emph{interface} associated to the configuration of the abstract component that it implements.

Let
\makebox{
\begin{scriptsize}
\begin{math}
C=\left[\makebox{\mbox{
\begin{minipage} {\textwidth}
\begin{tabbing}
$\mathit{kind}\ C\ [X_i:T_i\ ^{i=1{\dots}n}]\ (a_i\ \ ^{i=1{\dots}k})$ \\
$\mathbf{beg}$\=$\mathbf{in}$ \\
     \> $kind_i\ a_i : U_i\ (a_{r_{ij}}\ ^{j=1{\dots}p_i})\ \ ^{i=1{\dots}l}$\\
     \> $\mathbf{unit}\ u_i\ \ \ ^{i=1{\dots}q}$ \\
     \> $\mathbf{beg}$\=$\mathbf{in}$ \\
     \>      \>$\mathbf{slice}\ s_{ij}\ \mathbf{from}\ a_{z_{ij}}.v_{ij}\ \ ^{j=1{\dots}m_i}$ \\
     \>      \>$\mathbf{action}\ A_i$\\
     \> $\mathbf{end}$\\
$\mathbf{end}$
\end{tabbing}
\end{minipage}}
} \right]
\end{math}
\end{scriptsize}
}
be a configuration schema of an abstract component. $\mathcal{S}(C)$ maps to a tuple of C\# interfaces, one for each unit $u_i$, with the structure\footnote{Note about notation: \textcolor{blue}{$\left[\textcolor{black}{A_i}\ \ \textcolor{red}{\fbox{$P_i$}} \right]^{i=1 \dots n}$} means a sequence that include the elements $A_i$, for $i \in \{1,\dots,n\}$, such that predicate $P_i$ is valid.}

\begin{scriptsize}
\begin{tt}
\begin{minipage}{\textwidth}
\begin{tabbing}

namespace $C$ \\
\{\= \\
\>public \= interface $\mathbf{I}^{u_i}\left\langle\overline{\textbf{X}}\right\rangle : \mathbf{I}^{\mathit{kind}}$ \\
\>       \>   \hspace*{3mm} \makebox{
            \begin{math}\textcolor{blue}{
                \left[\makebox{\mbox{
                        \begin{minipage} {\textwidth}
                        \begin{tabbing}
                                \textcolor{black}{where $X_j:\mathbf{I}^{T_j.w_{ij}}$}
                        \end{tabbing}
                        \end{minipage}}}\ \ \textcolor{red}{\fbox{\rm a slice $T_j.w_{ij}$ exists (transitively) in $u_i$}}\ \textcolor{green}{\fbox{$X_j \in \overline{\textbf{X}}$}}
                \ \right]^{j=1 \dots n}
            }\end{math}}\\
\>\{ \= \\
\>   \>\ \makebox{
        \begin{math}\textcolor{blue}{
            \left[
                \makebox{\mbox{
                    \begin{minipage} {\textwidth}
                    \begin{tabbing}
                        \textcolor{black}{$\mathbf{I}^{U_{z_{ij}}.v_{ij}}$ $S_{ij}$ \{set;\};}
                    \end{tabbing}
                    \end{minipage}}}\ \ \textcolor{red}{\fbox{$z_{ij} \in \{1,\dots,k\}$}}
           \ \right]^{j=1 \dots m_i}
        }\end{math}
    }\\
%\> \>\\
\>\}\\
\}
\end{tabbing}
\end{minipage}
\end{tt}
\end{scriptsize}

The index $i$ refers to the unit ($u_i$). The interface declares a set of properties $S_{ij}$, one for each slice of the unit that corresponds to a unit of a public inner component. They require only their \emph{set} access method. The reason will be clarified in the next paragraphs. The notation used for abstracting C\# interface identifiers,  $\mathbf{I}^{U_{z_{ij}}.v_{ij}}$, means the name of the C\# interface that correspond to the unit $v_{ij}$ of the configuration  $U_{z_{ij}}$.

Let
\makebox{
\begin{scriptsize}
\begin{math}
CImpl=\left[\makebox{\mbox{
\begin{minipage} {\textwidth}
\begin{tabbing}
$\mathit{kind}\ $\=$C\ [X_i:T_i\ \ ^{i=1{\dots}n}]$\\
                 \>$\mathbf{implements}\ C'\ [X_i\ ^{i=1{\dots}n}]$ \\
$\mathbf{beg}$\=$\mathbf{in}$ \\
     \> $\mathbf{unit}\ u_i\ \ \ ^{i=1{\dots}q}$ \\
     \> $\mathbf{beg}$\=$\mathbf{in}$ \\
     \>      \>$\emph{\textbf{C}}_i$\\
     \> $\mathbf{end}$\\
$\mathbf{end}$
\end{tabbing}
\end{minipage}}
} \right]
\end{math}
\end{scriptsize}
} be a configuration of a \#-component. $\mathcal{S}(CImpl)$ maps to a set of C\# classes, one for each unit $u_i$, with the following structure:

\begin{scriptsize}
\begin{tt}
\begin{minipage}{\textwidth}
\begin{tabbing}
namespace $CImpl$ \\
\{\=\\
\>public \= class $\mathbf{H}^{u_i}\left\langle\overline{\textbf{X}}\right\rangle$: $\mathbf{Unit}$, $C.\mathbf{I}^{u_i}\left\langle\overline{\textbf{X}}\right\rangle$ \\
\>       \>   \hspace*{3mm} \makebox{
            \begin{math}\textcolor{blue}{
                \left[\makebox{\mbox{
                        \begin{minipage} {\textwidth}
                        \begin{tabbing}
                                \textcolor{black}{where $X_j:\mathbf{I}^{T_j.w_{ij}}$}
                        \end{tabbing}
                        \end{minipage}}}\ \ \textcolor{red}{\fbox{\rm a slice $T_j.w_{ij}$ exists (transitively) in $u_i$}}\ \textcolor{green}{\fbox{$X_j \in \overline{\textbf{X}}$}}
                \ \right]^{j=1 \dots n_i}
            }\end{math}}\\
\>\{ \= \\
\>   \> \textbf{\emph{// private slices}} \\
\>   \>\ \makebox{
        \begin{math}\textcolor{blue}{
            \left[
                \makebox{\mbox{\textcolor{black}{
                    \begin{minipage} {\textwidth}
                    \begin{tabbing}
                        private $\mathbf{I}^{U_{z_{ij}}.v_{ij}}$ $s_{ij}$ = null;\\
                        private $\mathbf{I}^{U_{z_{ij}}.v_{ij}}$ $S_{ij}$ \\
                        \{\= \\
                          \> set \=\{\ \= \\
                          \>     \>    \> $s_{ij} = value$;\\
                          \>     \>    \> \hspace*{8mm} \makebox{
                                            \begin{math}\textcolor{blue}{
                                                \left[
                                                    \makebox{\mbox{
                                                        \begin{minipage} {\textwidth}
                                                        \begin{tabbing}
                                                                \textcolor{black}{$s_{ik}.S'_{i{jk}} = value$;}
                                                        \end{tabbing}
                                                        \end{minipage}}}\ \textcolor{red}{\fbox{\rm $a_{z_{ik}}$ is a parameter of $a_{z_{ij}}$}}
                                                \ \right]^{k=1 \dots m_{i}}
                                            }\end{math}
                                          }\\
                          \>     \> \} \\
                        \}
                    \end{tabbing}
                    \end{minipage}}}} \ \textcolor{red}{\fbox{$z_{ij} \in \{k{+}1,\dots,l\}$}}
           \ \right]^{j = 1 \dots m_i}
        }\end{math}
    }\\
\>   \>\\
\>   \> \textbf{\emph{// public slices}} \\
\>   \>\ \makebox{
        \begin{math}\textcolor{blue}{
            \left[
                \makebox{\mbox{\textcolor{black}{
                    \begin{minipage} {\textwidth}
                    \begin{tabbing}
                        private $\mathbf{I}^{U_{z_{ij}}.v_{ij}}$ $s_{ij}$ = null;\\
                        public $\mathbf{I}^{U_{z_{ij}}.v_{ij}}$ $S_{ij}$ \\
                        \{\= \\
                          \> set \=\{\ \= \\
                          \>     \>    \> $s_{ij} = value$;\\
                          \>     \>    \> \hspace*{8mm} \makebox{
                                            \begin{math}\textcolor{blue}{
                                                \left[
                                                    \makebox{\mbox{
                                                        \begin{minipage} {\textwidth}
                                                        \begin{tabbing}
                                                                \textcolor{black}{$s_{ik}.S'_{i{jk}} = value$;}
                                                        \end{tabbing}
                                                        \end{minipage}}}\ \textcolor{red}{\fbox{\rm $a_{z_{ik}}$ is a parameter of $a_{z_{ij}}$}}
                                                \ \right]^{k=1 \dots m_{i}}
                                            }\end{math}
                                          }\\
                          \>     \> \} \\
                        \}
                    \end{tabbing}
                    \end{minipage}}}}
            \ \textcolor{red}{\fbox{$z_{ij} \in \{1,\dots,k\}$}}
           \ \right]^{j = 1 \dots m_i}
        }\end{math}
    }\\
\>   \>\\
\>   \> \textbf{\emph{// creation of private slices}}\\
\>   \> public void \underline{createSlices}()\\
\>   \> \{\ \=\\
\>   \>   \> base.createSlices();\\
\>   \>   \>\hspace*{3mm} \makebox{
            \begin{math}\textcolor{blue}{
                \left[\makebox{\mbox{
                        \begin{minipage} {\textwidth}%
                        \begin{tabbing}
                        \textcolor{black}{this.$S_{ij}$ = $\left(\mathbf{I}^{U_{z_{ij}}.v_{ij}}\right)$ \underline{BackEnd.createSlice}(this,$\dots$);}
                        \end{tabbing}%}
                        \end{minipage}}} \ \textcolor{red}{\fbox{$z_{ij} \in \{k{+}1,\dots,l\}$}}
                \ \right]^{j = 1 \dots m_i}
            }\end{math}}\\
\>   \> \}\\
\>  \> \\
\>   \> $\emph{\textbf{C}}_i$ \emph{// kind dependent part}\\
\> \}\\
\}
\end{tabbing}
\end{minipage}
\end{tt}
\end{scriptsize}

The properties associated to the \emph{public slices} of unit $u_i$, named $S_{ij}{^{\tiny\ |\ j \in \{1,\dots,m_i\} \wedge z_{ij} \in \{1,\dots,k\}}}$, required by interface $C.\mathbf{I}^{u_i}$, are implemented. In addition, there are private properties for the \emph{private slices} of $u_i$, named $S_{ij}{^{\tiny\ |\ j \in \{1,\dots,m_i\} \wedge z_{ij} \in \{k{+}1,\dots,l\}}} $.
%For each \emph{slice property} $S_{ij}{^{\tiny\ |\ j=1{\dots}m_i}}$, the writing access method (\emph{set}), besides to write the instance object in the field $s_{ij}$, writes to the public slices of other ....

The public method \textsf{createSlices} and the static method \textsf{BackEnd.createSlice} form a mutually recursive pair. When creating a unit $u_i$ of a \#-component $c$, \textsf{createSlices} calls \textsf{BackEnd.createSlice} to create the unit $v_{i{j}}$ of some \emph{private inner component} $a_{z_{i{j}}}$ of $c$ that is a private slice $s_{i{j}}$ of $u_i$. Then, after instantiating $v_{i{j}}$, \textsf{BackEnd.createSlice} calls \textsf{createSlices} for creating its slices. The procedure proceeds recursively until units with no slices are reached (primitive units). In the return, the object that represents $v_{i{j}}$ is assigned to the slice property $S_{i{j}}$, causing a call of its writing access method (\emph{set}). If $a_{z_{i{j}}}$ supplies any public inner component of another inner component $a_{z_{i{{k}}}}$, then $v_{ij}$ is also assigned to the corresponding public slice $S'_{ijk}$ of $s_{i{{k}}}$, such that $s_{i{{k}}}{\mapsto}a_{z_{i{{k}}}}.v_{i{{k}}} \in \sigma_i$.

Moreover, \textsf{BackEnd.createSlice} is a parallel method, since $q$ simultaneous calls are performed to create a \#-component $c$, each one executed by a process that has a unit $u_i\ ^{i=1{\dots}q}$ as slice. The $i^{th}$ call queries the DGAC (Distributed Global Assembly Cache), the HPE module responsible to manage parallel components, to find the class $\mathbf{H}^{u_i}$ that represent the unit $u_i$ of the best \#-component, deployed in the environment, for the abstract component referred by the inner component in the configuration. For that, DGAC uses the resolution algorithm of Figure \ref{fig:generalization_algorithm}.
%If none is found, a run-time exception is thrown. After instantiating an object from $H^{u_i}$, its method \textsf{createSlices} is called.

For each kind of \#-component, it may be defined a dependent part, referred as $\emph{\textbf{C}}_i$ in the schema. More specifically, $\emph{\textbf{C}}_i$ is the implementation of the interface defined by the interface $\mathbf{I}^{\mathit{kind}}$. For example, for the kind \emph{computation}, of HPE, it is defined the interface

\begin{center}
\begin{footnotesize}
\begin{tt}
\begin{minipage}{\textwidth}
\begin{tabbing}
inter\=face \textit{IComputation} \{ \\
     \> void compute();\\
\}
\end{tabbing}
\end{minipage}
\end{tt},
\end{footnotesize}
\end{center}

whose method \emph{compute} is implemented by the programmer to define the computation to be performed over the slices of each unit.

%\begin{figure}
%\centering
%\includegraphics[width=7cm]{figures/HPEExample_Simple_Configuration_2.eps}
%\caption{Visual Representation of the Configuration of the Application Example}
%\end{figure}

\subsection{Case Study}

 Figure \ref{fig:hierarchy_application_example} depicts the hierarchy of components of a configuration of an \emph{abstract component} of kind application for the parallel program of Section \ref{sec:hash_model}, named \textsc{AppExampleAbs}.
 The ellipses represent the transitive inner components that appear in the overall application.
 The arrows represent the ``is inner component of'' relation.
The colors assigned to the abstract components distinguish their kinds.
Dashed ellipses indicate parameters of the configuration, whose associated variable identifiers are italicized.

%\subsection{Implementation of Units in the Host Language}

\begin{figure}

\begin{tabular}{c|c}

\begin{minipage}{\textwidth}
\begin{tiny}
%\begin{verbatim}
\begin{tabbing}

01. namespace example.computation.\textsc{MatVecProduct} \\
02. \{ \= \\
03. \> public \=interface \=\emph{ICalculate}$<$C, E, N, Da, Dx, Dv$>$\\
04  \>        \>          \> : \emph{IComputationKind}\\
05. \> \>where C: \emph{ICluster}\\
06. \> \>where E: \emph{IEnvironment}$<$C$>$\\
07. \> \>where N: \emph{INumber}\\
08. \> \>where Da: \emph{IVecPartition}\\
09. \> \>where Dx: \emph{IVecPartition}\\
10. \> \>where Dv: \emph{IVecPartition}\\
11. \> \{\=\\
12. \>   \>	E \textbf{Env} \{set;\} \\
13. \>   \>	\emph{IParData}$<$C, E, \emph{Matrix}$<$N$>$, Da$>$ \textbf{A} \{set;\} \\
14. \>   \>	\emph{IParData}$<$C, E, \emph{Vector}$<$N$>$, Dx$>$ \textbf{X} \{set;\} \\
15. \>   \>	\emph{IParData}$<$C, E, \emph{Vector}$<$N$>$, Dv$>$ \textbf{V} \{set;\} \\
16. \> \}\\
17. \}

\end{tabbing}
%\end{verbatim}
\end{tiny}
\end{minipage}

&

\begin{minipage}{\textwidth}
\begin{tiny}
%\begin{verbatim}
\begin{tabbing}

01. namespace example.computation.impl.\textbf{\textsc{MatVecProductImplForNumber}} \{ \\
02. \ \ \ \ \= public \= class \textbf{\textit{HCalc}}\=\textbf{{\it ulate}}$<$C, E, N, Da, Dx, Dv$>$: \emph{Unit}, \emph{ICalculate}$<$C, E, D, Da, Dx, Dv$>$ \\
03. \> \> where C: \emph{IGNUCluster}\\
04. \> \> where E: \emph{IMPIFull}$<$C$>$\\
05. \> \> where N: \emph{INumber}\\
06. \> \> where Da: \emph{IByRows}\\
07. \> \> where Dx: \emph{IReplicate}\\
08. \> \> where Dv: \emph{IReplicate} \\
09. \> \{ \= \\
10. \> \> private E env = null;\\
11. \> \> private \emph{IParData}$<$C, E, \emph{Matrix}$<$N$>$, Da$>$ a = null;\\
12. \> \> private \emph{IParData}$<$C, E, \emph{Vector}$<$N$>$, Dx$>$ x = null;\\
13. \> \> private \emph{IParData}$<$C, E, \emph{Vector}$<$N$>$, Dv$>$ v = null;\\
14. \> \> \\
15. \> \> public \= E \textbf{Env} \{ set \{ \=this.env = a.Env = x.Env = v.Env = value; \} \}\\
16. \> \> public \= \emph{IParData}$<$C, E, \emph{Matrix}$<$N$>$, Da$>$ \textbf{A} \{ set \{ this.a = value; \} \}\\
17. \> \> public \= \emph{IParData}$<$C, E, \emph{Vector}$<$N$>$, Dx$>$ \textbf{X} \{ set \{ this.x = value; \} \} \\
18. \> \> public \= \emph{IParData}$<$C, E, \emph{Vector}$<$N$>$, Dv$>$ \textbf{V} \{ set \{ this.v = value; \} \}\\
19. \> \> \\
20. \> \> public \textbf{\textit{HCalculate}}() \{ $\cdots$ \} \\
21. \> \> public void createSlices() \{ $\cdots$ \} \\
22. \> \> public void \= compute() \{ \\
23. \> \> \> ($\cdots$)\\
24. \> \> \> \emph{IVector}$<$N$>$ arr = \textbf{V}.Value;\\
25. \> \> \> ($\cdots$)\\
26. \> \> \> // \emph{\textbf{\underline{1st attempt (unsafe). line 33 causes type check error} !!!}}\\
27. \> \> \> N newValue = new example.data.impl.NumberImpl.\textbf{\textit{INumberImpl}}(); \\
28. \> \> \> ($\cdots$)\\
29. \> \> \> // \emph{\textbf{\underline{2nd attempt (safe). In line 36, using reflection, an instance of N is created.}}}\\
30. \> \> \> N newValue = Activator.CreateInstance(typeof(N)); \\
31. \> \> \> ($\cdots$)\\
32. \> \> \> for \= (i=0; i<=arr.size(); i++) arr.set(i, newValue);\\
33. \> \> \> ($\cdots$)\\
34. \> \> \} \\
35. \> \}  \\
36. \}

\end{tabbing}
%\end{verbatim}
\end{tiny}
\end{minipage}

%\\
%(a) & (b)

\end{tabular}

\caption{Unit of \textsc{MatVecProduct} and $\textsf{MatVecProductImplForNumber}$}
\label{fig:vecvecproduct_units_code}
\end{figure}

The configuration of the inner component $\mathrm{V=A \times X}$ is \textsc{MatVecProduct}, discussed in Section \ref{sec:hcl_configuration}.
To illustrate how the proposed type system fits CTS (Common Type System), of CLI virtual machines, the interface $\texttt{ICalculate}$,
associated with the units of the abstract component $\textsc{MatVecProduct}$, and the class $\texttt{HCalculate}$,
associated with the units of $\textsf{MatVecProductImplForNumbers}$, are presented
in Figure \ref{fig:vecvecproduct_units_code}, obtained from the translation schemas introduced in the beginning of Section \ref{sec:implementation}.
$\textsf{MatVecProductImplForNumber}$ differs from $\textsf{MatVecProductImplForDouble}$ because it works with any number data type, including double precision float point ones.

It is important to understand how generic types of CLI are used to implement the relation between abstract components and their \#-components.
For instance, the interface $\texttt{ICalculate}$ is generic in type variables $C$, $E$, $N$, $Sa$, $Sx$,
and $Sv$, like $\textsc{MatVecProduct}$.
$\texttt{{HCalculate}}$ is also generic in the same type parameters, but
their bounds are specialized for the types for which the class is tuned, making possible to make assumptions about the structure of objects of these types. In the method \texttt{compute} of $\texttt{{HCalculate}}$ (lines 22 to 34), it is shown that CTS does
not allow that one instantiates
an object of class \texttt{INumberImpl}, implementing \texttt{INumber}, in
a context where a object of type $N$, such that $N<:\texttt{INumber}$, is expected, like in line 27 (1st attempt).
If the
value of $N$ at run-time is a proper subtype of $\texttt{INumber}$, like $\texttt{IDouble}$, the assignment to
array elements in line 32 is unsafe.
On the other hand, if the variable \emph{newValue} is instantiated like in line 30 (2nd attempt) it is created an object of
the actual type of $N$, which can be $\texttt{IDouble}$ safely.
This is the reason why languages such as C\# and Java
only support invariant generic types ($T\langle{U}\rangle <: T'\langle{U'}\rangle \Leftrightarrow T <: T' \wedge U = U'$).
In languages like
Java, where generic types are implemented using \emph{type erasure},
it is not possible to create an instance of the class associated to type variable $N$ at run-time, since
type variables are erased in compilation.
But this is possible in C\#, using reflection, because the bytecode of CIL (Common Intermediate Language) carries generic types at runtime. This is one of the motivations to use Mono for implementing HPE.

%This is another, yet less important, reason to adopt a CLI platform.

\section{Conclusions and Lines for Further Works}
\label{sec:conclusions}

The \# component model attempts to converge
software engineering techniques and parallel programming artifacts, addressing the raising in complexity and scale of recent applications in HPC domains.
The recent design and prototype of HPE, a \# programming
system, suggests gains in abstraction and modularity,
without significant performance penalties.
This paper introduced a type system for \# programming systems that
was applied to HPE, allowing the study of its formal properties,
mainly regarding safety, compositionability, and expressiveness. It has been designed for allowing programmers to make assumptions about specific features of parallel computing architectures, but also providing the ability to work at some desired level of abstraction. This is possible due to a combination of existential and universal bounded quantification.
In the near future, it is planned to research on the how other concepts found in higher-level type system designs may improve parallel programming practice.

\begin{footnotesize}
\bibliographystyle{sbc}
%\bibliography{../Language}

\begin{thebibliography}{}

\bibitem[{A}llan et~al. 2002]{Allan2002}
{A}llan, B.~A., {A}rmstrong, R.~C., {W}olfe, A.~P., {R}ay, J., {B}ernholdt,
  D.~E., and {K}ohl, J.~A. (2002).
\newblock {The CCA Core Specification in a Distributed Memory SPMD Framework}.
\newblock {\em {Concurrency and Computation: Practice and Experience}},
  14(5):{323--345}.

\bibitem[{A}lt et~al. 2004]{Alt2004}
{A}lt, M., {D}\"unnweber, J., {M}\"uller, J., and {G}orlatch, S. (2004).
\newblock {HOCs: Higher-Order Components for Grids}.
\newblock In {\em Workshop on Component Models and Systems for Grid
  Applications (in ICS'2004)}. Kluwer Academics.

\bibitem[{A}rmstrong et~al. 2006]{Armstrong2006}
{A}rmstrong, R., {K}umfert, G., {M}c{I}nnes, L.~C., {P}arker, S., {A}llan, B.,
  {S}ottile, M., {E}pperly, T., and {D}ahlgreen {T}amara (2006).
\newblock {The CCA Component Model For High-Performance Scientific Computing}.
\newblock {\em {Concurrency and Computation: Practice and Experience}},
  18(2):{215--229}.

\bibitem[{B}aduel et~al. 2007]{Baduel2007}
{B}aduel, L., {B}aude, F., and {C}aromel, D. (2007).
\newblock {Asynchronous Typed Object Groups for Grid Programming}.
\newblock {\em {Journal of Parallel Programming}}, 35(6):573--613.

\bibitem[{B}aude et~al. 2007]{Baude2007}
{B}aude, F., {C}aromel, D., {H}enrio, L., and {M}orel, M. (2007).
\newblock {Collective Interfaces for Distributed Components}.
\newblock In {\em {7th International Symposium on Cluster Computing and the
  Grid (CCGrid 07)}}. {IEEE Computer Society}.

\bibitem[{B}aude et~al. 2008]{Baude2008}
{B}aude, F., {C}aromel, F., {D}almasso, C., {D}anelutto, M., {G}etov, W.,
  {H}enrio, L., and {P}érez, C. (2008).
\newblock {GCM: A Grid Extension to Fractal for Autonomous Distributed
  Components}.
\newblock {\em {Annals of Telecommunications}}, 0:000--000.

\bibitem[{B}ernholdt {D}.~{E}. et~al. 2004]{Bernholdt2004}
{B}ernholdt {D}.~{E}., {N}ieplocha, J., and {S}adayappan, P. (2004).
\newblock {Raising Level of Programming Abstraction in Scalable Programming
  Models}.
\newblock In {\em {IEEE International Conference on High Performance Computer
  Architecture (HPCA), Workshop on Productivity and Performance in High-End
  Computing (P-PHEC)}}, pages {76--84}. {Madrid, Spain}, {IEEE Computer
  Society}.

\bibitem[{B}runeton et~al. 2002]{Bruneton2002}
{B}runeton, E., {C}oupaye, T., and {S}tefani, J.~B. (2002).
\newblock {Recursive and Dynamic Software Composition with Sharing}.
\newblock In {\em {European Conference on Object Oriented Programming
  (ECOOP'2002)}}. {Springer}.

\bibitem[{C}arvalho {J}unior et~al. 2007]{Carvalho2007a}
{C}arvalho {J}unior, F.~H., {L}ins, R., {C}orrea, R.~C., and {A}ra\'ujo, G.~A.
  (2007).
\newblock {Towards an Architecture for Component-Oriented Parallel
  Programming}.
\newblock {\em {Concurrency and Computation: Practice and Experience}},
  19(5):{697--719}.
\newblock Special Issue: Component and Framework Technology in High-Performance
  and Scientific Computing. Edited by David E. Bernholdt.

\bibitem[{C}arvalho {J}unior and {L}ins 2008]{Carvalho2008}
{C}arvalho {J}unior, F.~H. and {L}ins, R.~D. (2008).
\newblock {An Institutional Theory for \#-Components}.
\newblock {\em {Electronic Notes in Theoretical Computer Science}},
  195:113--132.

\bibitem[{C}arvalho {J}unior and {L}ins 2009]{Carvalho2009a}
{C}arvalho {J}unior, F.~H. and {L}ins, R.~D. (2009).
\newblock {Compositional Specification of Parallel Programs Using Circus}.
\newblock {\em {Electronic Notes in Theoretical Computer Science}}, 0000:0--0.
\newblock {5th International Workshop on Formal Aspects of Component Software}.

\bibitem[{D}ongarra et~al. 1996]{Dongarra96}
{D}ongarra, J., {O}tto, S.~W., {S}nir, M., and {W}alker, D. (1996).
\newblock {A Message Passing Standard for MPP and Workstation}.
\newblock {\em {Communications of ACM}}, 39(7):84--90.

\bibitem[{M}illi et~al. 2004]{Mili2004}
{M}illi, H., {E}lkharraz, A., and {M}cheick, H. (2004).
\newblock {Understanding Separation of Concerns}.
\newblock In {\em {Workshop on Early Aspects - Aspect Oriented Software
  Development (AOSD'04)}}, pages 411--428.

\bibitem[{P}ierce 2002]{Pierce2002}
{P}ierce, B. (2002).
\newblock {\em {Types and Programming Languages}}.
\newblock {The MIT Press}.

\bibitem[{P}ost and {V}otta 2005]{Post2005a}
{P}ost, D.~E. and {V}otta, L.~G. (2005).
\newblock {Computational Science Demands a New Paradigm}.
\newblock {\em {Physics Today}}, 58(1):35--41.

\bibitem[{S}arkar et~al. 2004]{Sarkar2004}
{S}arkar, V., {W}illiams, C., and {E}bcio$\breve{g}$lu, K. (2004).
\newblock {Application Development Productivity Challenges for High-End
  Computing}.
\newblock In {\em {IEEE International Conference on High Performance Computer
  Architecture (HPCA), Workshop on Productivity and Performance in High-End
  Computing}}, pages {14--18}.

\bibitem[van~der {S}teen 2006]{Steen2006}
van~der {S}teen, A.~J. (2006).
\newblock {Issues in Computational Frameworks}.
\newblock {\em {Concurrency and Computation: Practice and Experience}},
  18(2):{141--150}.

\bibitem[{W}ang and {Q}ian 2005]{Wang2005}
{W}ang, A. J.~A. and {Q}ian, K. (2005).
\newblock {\em {Component-Oriented Programming}}.
\newblock {Wiley-Interscience}.

\end{thebibliography}

\end{footnotesize}

\end{document}